# Measuring Poisson's ratio during nanoindentation via lateral contact stiffness – issues of slip and plasticity

Owen Brazil[1], George M. Pharr[1]

[1]*Department of Materials Science & Engineering, Texas A&M University, College Station, TX 77843-3003, USA*

*Correspondence to: brazilo@tcd.ie

# Abstract

Although Poisson's ratio, $\nu$, is a quantity of great importance in materials design, presently, very few methods are available to measure it at small scales (micrometre) and those typically involve intensive geometric refinement of the sample. In this work we show how Poisson's ratio may be measured during nanoindentation experiments by means of a second load actuator oriented orthogonally to the primary vertical loading axis. We apply a small oscillating lateral load to Berkovich and sphero-conical indenter tips during nanoindentation experiments into fused silica and metallic samples. This enables the elastic lateral contact stiffness to be measured as a function of depth in a manner analogous to continuous stiffness measurements in conventional nanoindentation experiments. This stiffness may in turn be used to calculate $\nu$. During constant strain rate experiments, we extract values of $\nu = 0.16$ for fused silica and $\nu = 0.34$ for polycrystalline aluminium over indentation depths of several hundred nanometres. We highlight two critical issues that must be addressed to ensure accurate measurement of $\nu$ during indentation: partial slip and additional plasticity brought about through the added shear under tangential loading. We study these separate phenomena in fused silica and single crystal nickel, suggesting that interfacial slip is favoured in harder, brittle materials, while plasticity is dominant in more ductile materials. We offer some suggestions on how to mitigate these problems based on tip geometry and loading protocols.

# Main Text

## 1. Introduction

Over the past 30 years, nanoindentation has become a key technology in elucidating the behaviour of matter at the small scales.[1], [2] Aided by advances in instrumentation and microfabrication technologies, it is now possible to directly access stress – strain, high strain rate, and elevated temperature data for a wide range of materials and tip geometries.[3]–[6] As well as informing the design of modern materials, nanoindentation has also revealed phenomena of fundamental scientific interest such as pressure induced phase transformations and intrinsic and extrinsic size effects.[7]–[9] Key to the widespread proliferation of indentation and enabling many of the these scientific breakthroughs has been the ability monitor properties like hardness and modulus continuously during loading via the continuous stiffness method (CSM), where a small oscillatory displacement or load is superposed onto the quasi-static signal, meaning the sample is continuously partially unloaded, thereby allowing the Oliver-Pharr relations to be applied throughout load-up.[10] The inclusion of a CSM signal supplied by means of a lock-in amplifier also allows more advanced sample/contact characterisation, e.g., quantification of creep and contact ageing during a hold segment or measurement,[11] or measurement of viscoelastic properties like storage and loss moduli through CSM frequency sweeps.[12] A long standing goal of experimentalists has been to extend the range and flexibility offered by normal CSM from one dimension to two by means of a second actuator and lock-in amplifier connected to the tip orthogonal to the primary vertical loading direction. In addition to enabling the study of interface and frictional properties,[13]–[15] *lateral* continuous stiffness measurement (LCSM) allows Poisson's ratio to be measured during indentation.[16] For isotropic materials, this enables total elastic characterisation of the sample and removes one of the more prominent assumptions of the Oliver-Pharr method for measuring hardness and elastic modulus by nanoindentation methods, i.e., that Poisson's ratio of the sample is known at least approximately prior to indentation.[2] Poisson's ratio is a quantity of fundamental scientific importance,[17] and may yield insights on packing density,[18] connectivity,[19] phase transformations,[20] and inelastic processes.[21] The ability to reliably measure $\nu$ at small scales represents a significant step forward in our ability to profile new materials at the nanoscale.

While some prototypical systems have been developed in the past,[14], [16] it is only recently that a commercially available dual-axis indenter capable of applying precise loads and displacements in two directions has been developed in the form of the Gemini 2D Multi-axis Transductor (Nanomechanics Inc, Oak Ridge, TN, USA). This system is composed of two 50 mN capacity load actuators mounted at 90 degrees to one another and connected to the indenter tip by means of two glass slides of approximately 1.5 cm length and 0.5 cm width and a special tip mounting block at their intersection. Quasi-static loads of up to 50 mN may be applied to the sample in either axis, with load and displacement resolutions in the nN and pm ranges. Each actuator has a dedicated lock-in amplifier enabling independent CSM and LCSM signals to be applied simultaneously. Letting $P_z$ be the indentation load, Poisson's ratio may be measured using the Gemini system via a standard constant $\dot{P}_z/P_z$ indentation experiment where CSM & LCSM signals are simultaneously superposed onto the quasi-static vertical load/displacement ($P_z/\delta_z$) signal. If a geometrically self-similar indenter like the Berkovich three-sided pyramid is used and the hardness is independent of indentation depth, this results in a constant indentation strain rate, $\dot{\varepsilon}_i$. The contact geometry is illustrated in Fig. 1a. Here, $\overline{p_z}$ is the vertical oscillating CSM load applied at 129 Hz whose peak-to-peak RMS amplitude is 5% of the quasi-static load at all times. Similarly, $\overline{q_x}$ is the LCSM load amplitude, applied at 100 Hz and equal to 2% of the quasi-static normal load during indentation. Note that all oscillation amplitudes used in this work are peak-to-peak RMS values. The two off resonance frequencies were chosen to be non-integer multiples of one another to limit mechanical cross-talk at the contact. $\overline{\delta_z}$ and $\overline{\delta_x}$ are the respective resultant CSM displacement amplitudes. The measured vertical contact stiffness $S_z$ is related to the elastic properties of the contact via the familiar expression:

$$S_z = 2E^* \sqrt{\frac{A_c}{\pi}}, \qquad (1)$$

where $E^* = \left(\frac{1-v_1^2}{E_1} + \frac{1-v_2^2}{E_2}\right)^{-1}$ is the reduced elastic modulus, $E_n$ and $v_n$ the Young's modulus and Poisson's ratio of the tip and sample, and $A_c$ is the projected contact area[22]. Following suitable frame stiffness calibration, the lateral contact stiffness, $S_x$, may be written in the form originally developed by Mindlin[23]:

$$S_x = 8G^* \sqrt{\frac{A_c}{\pi}}, \qquad (2)$$

where $G^* = \left(\frac{1-2v_1}{G_1} + \frac{1-2v_2}{G_2}\right)^{-1}$ is the reduced shear modulus. These expressions may be combined to give the Poisson's ratio $v$ as[24]:

$$v = \frac{2\left(1 - S_x/S_z\right)}{\left(2 - S_x/S_z\right)}. \qquad (3)$$

Fig. 1b shows the normal load – displacement curves for indentation into fused silica (blue) and polycrystalline aluminium (orange) samples taken with a Berkovich tip at $\dot{\varepsilon}_l = 0.05\ s^{-1}$. The measured $S_z$ and $S_x$ for the two materials are plotted as a function of indentation depth in Fig. 1c, where the observed linear behaviour is in keeping with equations 1 & 2 as $\sqrt{A_c} \propto \delta_z$ for a geometrically self-similar Berkovich tip. The Poisson's ratios calculated via equation 3 are plotted in Fig. 1d, with reasonable values of $v = 0.16$ for fused silica and $v = 0.34$ for aluminium extracted once $\nu$ converges to near constant values at approximately $\delta_z = 150$ nm and $\delta_z = 200\ nm$ respectively.[25] Given that the samples required no special preparation or surface treatment beyond what is needed for standard indentation, figure 1 highlights the simplicity and practical advantage of the LCSM-based $\nu$ measurement method for nanometre scale volumes over existing techniques, such as where significant geometric refinement of the sample is required.[26]–[28]

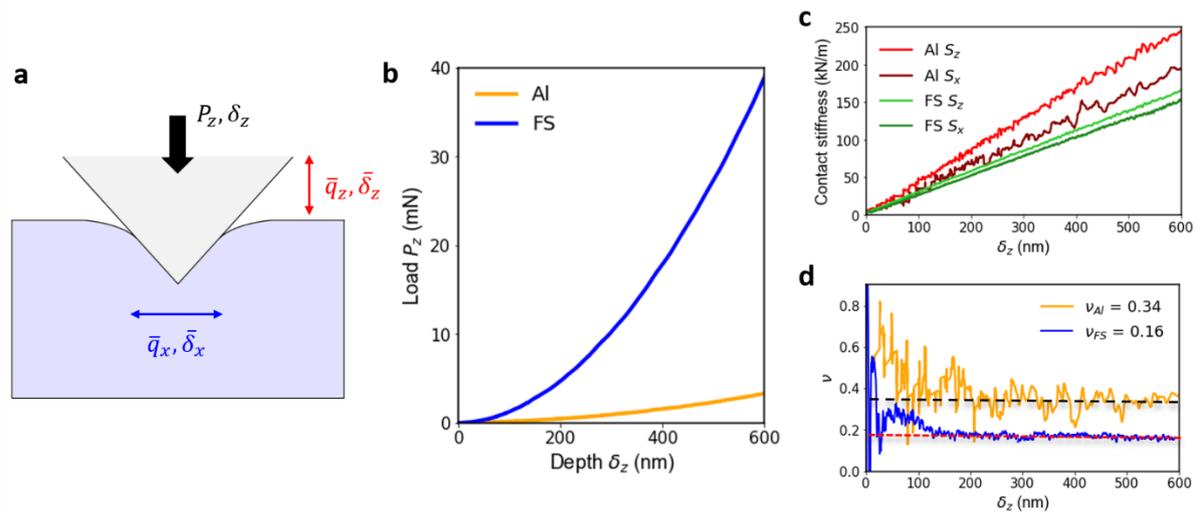

**Figure 1.** (a) Schematic of contact between a pyramidal tip and sample under CSM and LCSM loads. (b) Load – displacement curves for Berkovich indentation into fused silica and polycrystalline aluminium. (c) Corresponding normal and lateral contact stiffnesses. (d) Poisson's ratio as a function of depth via equation 3.

A central caveat of CSM testing is that the addition of the small oscillating load/displacement should not measurably affect the behaviour of the contact, in particular the stiffness.[29] However, even in the case of a standard one dimensional indentation experiment, CSM loading may introduce significant complications to the analysis of sample mechanical properties, potentially leading to erroneous measures of hardness and modulus. For example, two sources of potential error have been identified with vertical CSM measurements: (1) a lower than expected stiffness at low indentation depths due to preferential unloading in high modulus to hardness materials,[30], [31] and (2) a 'plasticity error' at large depths and low CSM frequencies where the contact spends a significant portion of the CSM loading/unloading cycle deforming plastically, in direct contradiction to the assumption that the oscillation is predominantly elastic.[32]–[34] Similar considerations exist for LCSM measurements. In addition, the validity of equation 2 rests upon two questionable assumptions: (1) the contact is in a state of full stick, and (2) the addition of the lateral load causes no further plasticity beyond what would be present under purely quasi-static loading.

In this work we address both of these assumptions. First, we consider the influence of partial slip and the stick – slip transition on the lateral contact stiffness of fused silica indented with a Berkovich tip. We then show how additional plasticity causes a less stiff response in single crystal nickel indented with a 1 μm radius spherical tip when the lateral oscillating load $\overline{q_x}$ is a significant proportion of the quasi-static load. Finally, we suggest that as with the aforementioned normal CSM errors, these LCSM issues can be mitigated by limiting the $\overline{q_x}$ to a small percentage of the applied quasi-static normal load.

## 2. Materials and methods

All indentation experiments considered here are performed using the Gemini dual-axis nanoindentation system. The lateral frame stiffness, $S_f$, of the system is determined as 105,000 N/m. The lateral contact stiffness, $S_x$, is then determined from the measured raw lateral stiffness, $S_{raw}$, by treating the frame and contact as linear spring elements in series: $S_x^{-1} = S_{raw}^{-1} - S_f^{-1}$. All indents are performed at constant $\frac{\dot{P}_z}{P_z} = 0.1\ s^{-1}$, which for the Berkovich tip used to measure $\nu$ in the introductory section and to study partial and gross slip in section 3 corresponds to a constant indentation strain rate of $\dot{\varepsilon}_i = 0.05\ s^{-1}$. Measurement of quantities derived from the normal contact stiffness $S_z$ such as elastic modulus $E$, hardness $H$, contact area $A_c$, and contact radius $a$ is accomplished via normal (vertical) CSM oscillation in line with conventional nanoindentation best practices. For all LCSM experiments, the lateral oscillation is applied at a frequency of 100 Hz, and the normal CSM is applied at 129 Hz. The fused silica and polycrystalline aluminium samples were supplied as reference materials accompanying the nanoindenter via Nanomechanics Inc. The single crystal nickel sample was purchased from Surface Preparation Laboratory, Zandaam, NL and mechanically polished to a mean roughness of less than 5 nm.

## 3. Partial and gross slip in fused silica

As was mentioned in the first section, equation 2 assumes that no slip occurs at the contact between the tip and sample. Essentially, this state of 'full-stick' means there is no relative motion between coincident points on the surfaces directly at the interface, while far-field motion in the respective bodies may still occur.[35] The other extreme, gross slip, allows relative motion of initially coincident points across the entire surface, albeit at some energetic cost due to frictional resistance. For a single asperity contact, the transition between stick and slip was formulated by Bowden & Tabor to take place at a critical lateral load, $Q_{slip}$, dependent on the contact area and the adhesive shear strength of the interface $\tau_0$[36] according to:

$$Q_{slip} = \tau_0 A_c \qquad (4)$$

While some divergence from equation 4 is encountered at very small scales where the discrete effects of atomic scale roughness emerge,[37], [38] it is generally thought applicable to contacts with radii of several hundred nm, as is typical in indentation experiments. Within this framework, it should be possible to design a LCSM protocol where slip has minimal influence on the lateral stiffness. For a perfect Berkovich tip and a material like fused silica in which the hardness is independent of depth, the contact area $A_c$ is a linear function of $P_z$ during load-up.[2] Therefore, by specifying that the lateral load $\overline{q_x}$ be a constant percentage of $P_z$ during indentation, the ratio $\overline{q_x}/A_c$ is also constant throughout. As such, if:

$$\frac{\overline{q_x}}{A_c} < \frac{\tau_0}{\sqrt{2}} \qquad (5)$$

the contact should remain in a high stiffness, full-stick state and equation 2 should be applicable. We note the $\sqrt{2}$ in equation (5) is a consequence of $\overline{q_x}$ being a peak-to-peak RMS value. Equation 5 is readily testable by performing a series of constant strain rate indentations at $\dot{\varepsilon}_i = 0.05\ s^{-1}$ with a Berkovich tip into fused silica, applying LCSM, and using different values of $\overline{q_x}$. This loading procedure is shown in Fig. 2a, with the inset showing five $\overline{q_x}$ values from 1-5% of $P_z$. The corresponding measured lateral contact stiffnesses $S_x$ are plotted as functions of depth in Fig. 2b. While the expected linear behaviour is seen for the 1% and 2% LCSM indents, a distinct transition from a state of high stiffness to low is observed at higher percentages, at odds with the description above. This transition becomes more abrupt and occurs closer to the initial surface (smaller depths) as the LCSM load percentage is increased. Fig. 2c shows that this transition corresponds to a large, sudden increase in the lateral oscillation displacement $\overline{\delta_x}$, which we assert to be a shift to a state of gross slip across the whole contact. The state of relative low stiffness and high oscillation amplitude essentially persists through the remainder of the indentation experiments, with the contact never recovering to the projected Mindlin stiffness (dashed black line).

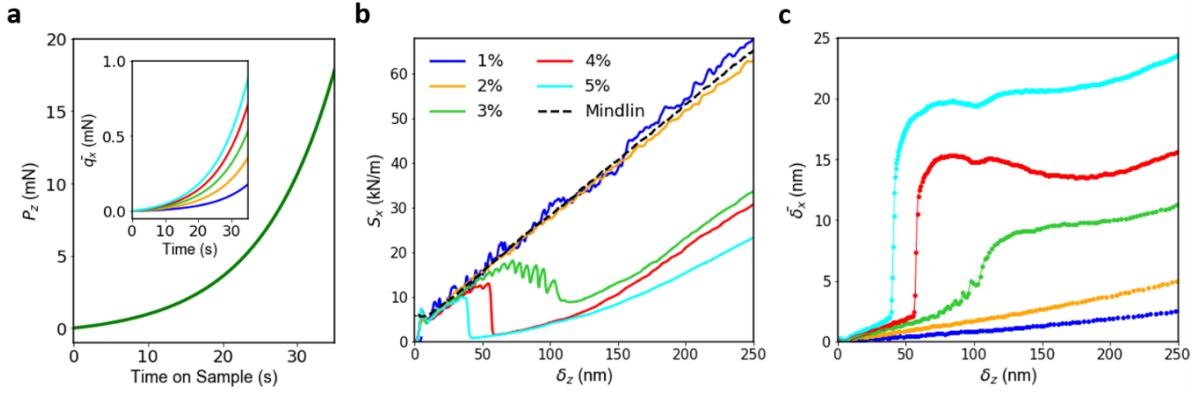

**Figure 2.** (a) Loading procedure used in constant strain rate LCSM experiments in fused silica with a sharp Berkovich indenter tip. Inset shows RMS lateral oscillating load for 5 experiments, where $\overline{q_x}$ is a fixed percentage of $P_z$ ranging from 1 – 5%. (b) Measured lateral contact stiffnesses as a function of indentation depth. Mindlin solution is plotted via equation 2 in black. (c) RMS lateral tip displacement versus depth as measured via the lock-in amplified. The drops in stiffness in (b) correspond to a sudden increase in lateral displacement.

Evidently something is missing in the above analysis to allow for this transition to a slip dominated friction state from the initial one conforming to the elastic solution. To gain additional insight, it is useful to consider the shear stress distribution at the contact interface $\tau_{x,z=0}(r)$ for an axisymmetric elastic contact of radius $a$ subjected to simultaneous normal and tangential loads[35]:

$$\tau_{x,z=0}(r) = \frac{Q_x}{2\pi a\sqrt{a^2 - r^2}} \, , \tag{6}$$

where $Q_x$ is the applied lateral load. In order to relax the stress singularities existing at the contact edge $r = a$ in equation 6, Mindlin proposed that the outer regions of the contact could slip while the inner regions remained in a state of stick[23]. As the magnitude of the lateral load is increased, the size of the slip region, an annulus of radius $c$ extending from the contact edge inward, also increases[39]. This scenario is sketched out Fig. 3a, where we have replaced $Q_x$ with the LCSM load amplitude $\overline{q_x}$ appropriate to our experiment. This state of partial slip (sometimes called micro-slip) was originally formulated for macroscopic contact pairs, where the ratio of tangential to normal forces $\mu = Q_x/P_z$ is fixed in the slip zone at the macroscopic coefficient of sliding friction[35], [39]. Recently, we have shown that this description is equally applicable to single asperity micro-contacts, where instead of fixed $\mu$ in the slip region, the shear stress at the interface in this region is fixed at the $\tau_0$[15]. This substitution is reflective of the fact that for a single asperity contact, the apparent contact area is essentially equal to the real contact area, whereas at macroscopic scales surface roughness limits real contact to a relatively small percentage of the apparent area[40].

Extending the work of *Lucas* and *Gao*[13], [24], we were able to demonstrate that partial slip occurs for both elastic and plastic diamond-silica contacts, formed with both spherical and Berkovich indenter tips. In this configuration, the tangential stress distribution in the stick region at the centre of the contact is modified to:

$$\tau_{x,z=0}(r) = \frac{\tau_0}{\pi} \cos^{-1}\left[\frac{2c^2 - a^2 - r^2}{a^2 - r^2}\right], \quad r \leq c \ . \tag{7}$$

Meanwhile, in the slipping annulus the tangential stress is fixed at $\tau_0$. This modification of contact stress leads to a new expression for lateral contact stiffness $S_x$ dependent on the size of the slip annulus, $c$, and therefore the magnitude of the lateral oscillation, given by:

$$S_x^{slip} = 8cG^* = 8aG^* \sqrt{1 - \left(\frac{2G^*}{\tau_0 a}\right)^2 \overline{\delta_x^2}} \ . \tag{8}$$

The key results of that study are reproduced in Fig. 3b & 3c, for which a 5 μm radius spherical indenter tip has been used instead of the Berkovich pyramid, and a fixed normal load of 3 mN is applied to the spherical tip to make contact with the fused silica sample. The lateral oscillation magnitude is slowly ramped up, with the resultant RMS load versus displacement plot shown in Fig. 3b. Equations 2 and 8 are plotted as red and blue lines respectively, where a interfacial shear strength of $\tau_0 = 0.3$ GPa is utilized. The contact stiffness is plotted as a function of $\overline{\delta_x}$ in Fig. 3c, where a gradual decrease in $S_x$ occurring in accordance with equation 8 until $\overline{\delta_x} \sim 2.4$ nm is observed, at which point the slip annulus extends all the way to $r = 0$ and the contact enters a state of gross slip, with $S_x$ rapidly dropping to near zero. The model of partial slip was found to work well at a series of indentation depths with a constant $\tau_0$. Further, while originally formulated for purely elastic contacts, we demonstrated that this description could be applied to contacts deforming plastically under normal contact pressures. This is particularly relevant to the case of constant strain rate indentation, where for indenter geometries like Vickers or Berkovich pyramidal tips, the sample is typically in a state of fully developed plasticity throughout loading.

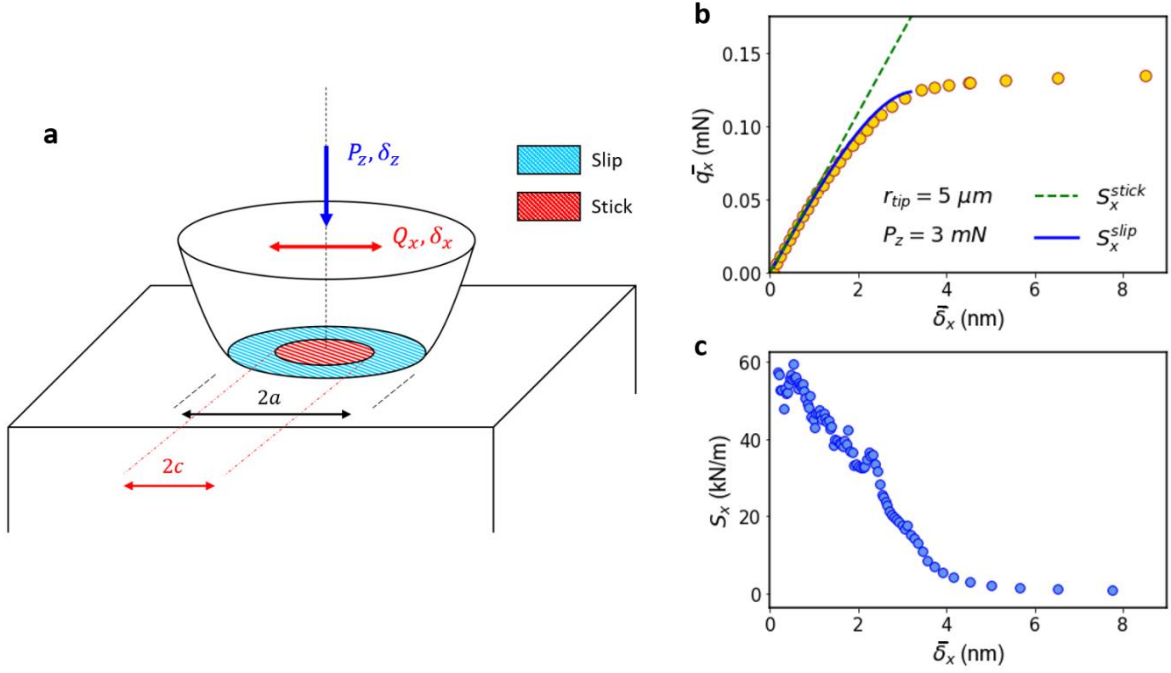

**Figure 3.** (a) Schematic of the partial interface slip model. Stress concentrators at the contact edge are relieved by slip occurring in an annulus of internal radius $c$. (b) RMS lateral load – displacement curve for a contact formed of an elastic contact formed of a 5 μm radius spherical indenter and a fused silica sample. Equation 2 is plotted in red and equation 8 in blue. (c) Measured lateral contact stiffness as a function of the magnitude of lateral oscillation.

We assert that the transitions to a gross slip state seen in figure 2 during constant strain rate indentation are facilitated by partial slip at the contact edge, in a manner in keeping with the description supplied above. The principal difference between the previous study and the current set of experiments is that in the present case, the contact radius $a = \sqrt{A_c/\pi}$ is continuously increasing as $P_z$ is increased during indentation. For an idealised spherical or axisymmetric tip, this sets up an interesting dynamic between $a$ and $\overline{\delta_x}$, as the rates at which they increase would determine whether or not the contact transition from a state of partial slip to gross slip or adheres more closely to the Mindlin solution. Equation 8 may be rewritten in the form:

$$S_x^{slip} = 8G^* \sqrt{a^2 - \left(\frac{2G^*}{\tau_0}\right)^2 \overline{\delta_x}^2} \tag{9}$$

which implies the contact will slip if a lateral oscillation magnitude of:

$$\overline{\delta_{x_{slip}}} = \frac{1}{2}\left(\frac{\tau_0}{G^*}\right)a \qquad (10)$$

is achieved. Based on our previous research and similar studies with pyramidal tip geometries,[14], [15] we assume the above formulation can be extended from the case of a spherical tip to that of a pyramidal Berkovich tip in manner similar to how a such tips can be approximated as axisymmetric conical indenters. Due to the self-similarity of the Berkovich tip used here, $a$ increases linearly with the tip depth $\delta_z$, with $a \sim \sqrt{24.5/\pi}\,\delta_z$ for an ideally sharp tip[41]. If it is assumed that the LCSM displacement depends linearly on the LCSM load, an assumption which should be valid for small oscillations[42], then $\overline{\delta_x} \propto P_z$, the normal load. From Fig. 1b, it is apparent that $P_z$ increases in a power law fashion with increasing depth $\delta_z$ for a Berkovich tip during load-up. As such, the LCSM amplitude $\overline{\delta_x}$ will eventually grow faster than $a$ as the tip penetrates the surface, and at some depth the condition $\overline{\delta_x} = \overline{\delta_{x_{slip}}}$ will be achieved. The depth at which this occurs will depend on the relative magnitude of $\overline{q_x}$, with larger $\overline{q_x}$ generating a larger lateral displacement, and as such resulting in earlier slip, as is seen in Figs. 2b & 2c. According to the partial slip model outlined here, the transition to gross slip should occur at a fixed $\overline{\delta_x}/a$ for all tests. In Fig. 4a, this quantity is plotted as a function of depth for the Berkovich tip experiments conducted in Fig. 2, where $a$ has been derived from the normal CSM signal applied to the contact. It is indeed observed that gross slip is achieved at $\overline{\delta_x}/a = 0.017$ for the $\overline{q_x} = 1 - 3\%\, P_z$ indents. This value is marked by the dashed grey line in Fig. 4a and, referring to equation 10, gives a interface shear strength to reduced shear modulus ratio of $\tau_0/G^* \sim 1/45$ once it is taken into account that $\overline{\delta_x}$ is an RMS value. This measurement is in line with other studies measuring the interfacial shear strength of diamond – silica contacts[13], [43], supporting our assertion that the dramatic drop in stiffness corresponds to a transition from partial to gross slip. With this quantity determined, it is possible to calculate the critical magnitude of lateral displacement required to induce gross slip, $\overline{\delta_{x_{slip}}}$, as a function of depth $\delta_z$ via equation 10. This is plotted as the black line in Fig. 4b, alongside the 5 LCSM indentation experiments considered in Fig. 4a. In this case the values for $a$ used to obtain $\overline{\delta_{x_{slip}}}$ are measured in a separate indentation experiment where only normal (vertical) CSM is applied. This procedure was adopted as an additional safeguard to ensure the superposition of LCSM and CSM versus CSM did not affect the vertical contact stiffness relations. We note that as a result of this methodology, $\overline{\delta_{x_{slip}}}$ does not cross the y-axis at the origin, but rather a finite value due to $a$ being derived from the vertical contact stiffness, which rises rapidly at the very early stages of indentation. Despite this, the line intersects the 3 – 5% indents at the point at which $\overline{\delta_x}$ dramatically increases due to slip, as predicted. The 1% and 2% indents do not intersect $\overline{\delta_{x_{slip}}}$ over the range considered here, thereby explaining their continuous behaviour throughout load-up. However, the slowly rising non-linear behaviour of these curves at greater depths as plotted in Fig. 4c suggests these tests too would eventually slip at sufficiently high $\delta_z$. The required normal loads $P_z$ and LCSM load to achieve these transitions are beyond the capacity of the Gemini system used here, however.

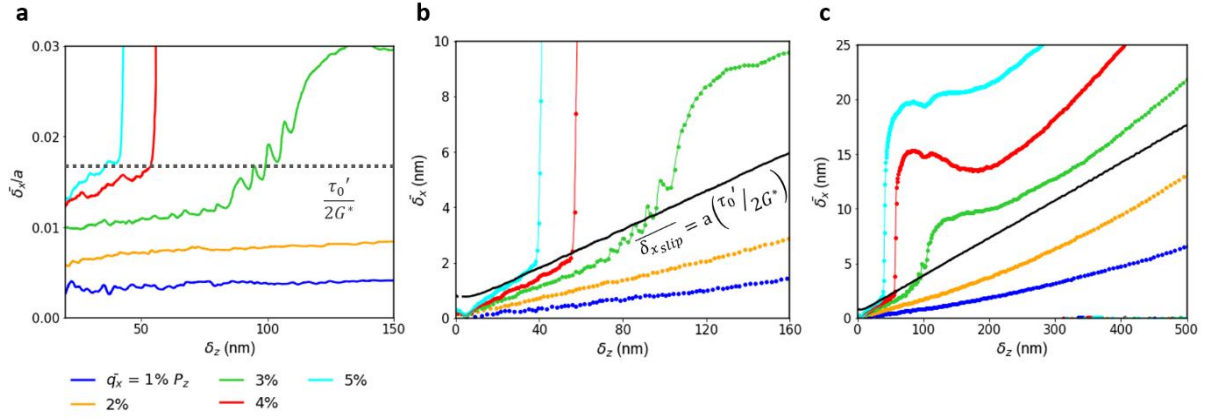

**Figure 4.** (a) Lateral oscillation displacement amplitude normalised over contact radius for the 5 LCSM indentation experiments into fused silica with a Berkovich tip. Dashed line shows the fixed ratio at which the contact fails for the 3 – 5% tests. (b) Critical lateral oscillation to induce slip $\overline{\delta_{x\,slip}}$ as a function of depth into surface. Plot is superimposed onto LCSM $\overline{\delta_x}$ curves and shows that failure occurs when $\overline{\delta_x} = \overline{\delta_{x\,slip}}$. (c) Replot of (b) with larger field of view.

It is clear from this analysis that any experiment seeking to measure Poisson's ratio in fused silica or comparable low modulus to hardness ratio materials like quartz or sapphire must use a low enough value of $\overline{q_x}$ to avoid the partial slip to gross slip transition. While some partial slip must occur to regularise stress singularities at the contact edge, the close adherence of the 1% and 2% $\overline{q_x}$ indents in this section indicate that if the LCSM loads employed are small enough, a good approximation to the elastic, full-stick solution may still be achieved. We conclude this section by noting that in spite of the clear distinction between the two states in terms of the lateral stiffness, the stick/slip state of the contact appears to have minimal impact on the normal stiffness and conventional $P_z - \delta_z$ load – displacement curve. This is shown in Fig. 5a where all five LCSM experiment curves are plotted alongside a standard constant strain rate indent into fused silica with no LCSM. There is no evidence of the dramatic changes in lateral dynamics leaking into the vertical quasi-static behaviour, with the close-up of the low load – displacement behaviour in Fig. 5b showing the 3 – 5% tests lie directly on top of the 1%, 2% and no LCSM curves for which the gross slip transition does not occur. An additional experiment shown in Fig. 5c where vertical CSM and LCSM are applied simultaneously (similar to in Fig. 1) further demonstrates the vertical contact stiffness $S_z$ is unaffected by the gross slip transition in $S_x$. Given that much of the mathematical framework supporting instrumented indentation was developed without explicit reference to friction or under the assumption of frictionless contact,[22], [44]–[46] the results of Fig. 5 are rather reassuring as they indicate that $S_z$ dependant quantities such as hardness and modulus may not be overly affected by the friction state of the contact. This point may not be generalisable, however, with a series of finite element studies suggesting tip/sample friction may have a significant bearing on the derived properties of materials that display significant pile-up or strain hardening and will depend on the tip geometry employed.[47]–[49]

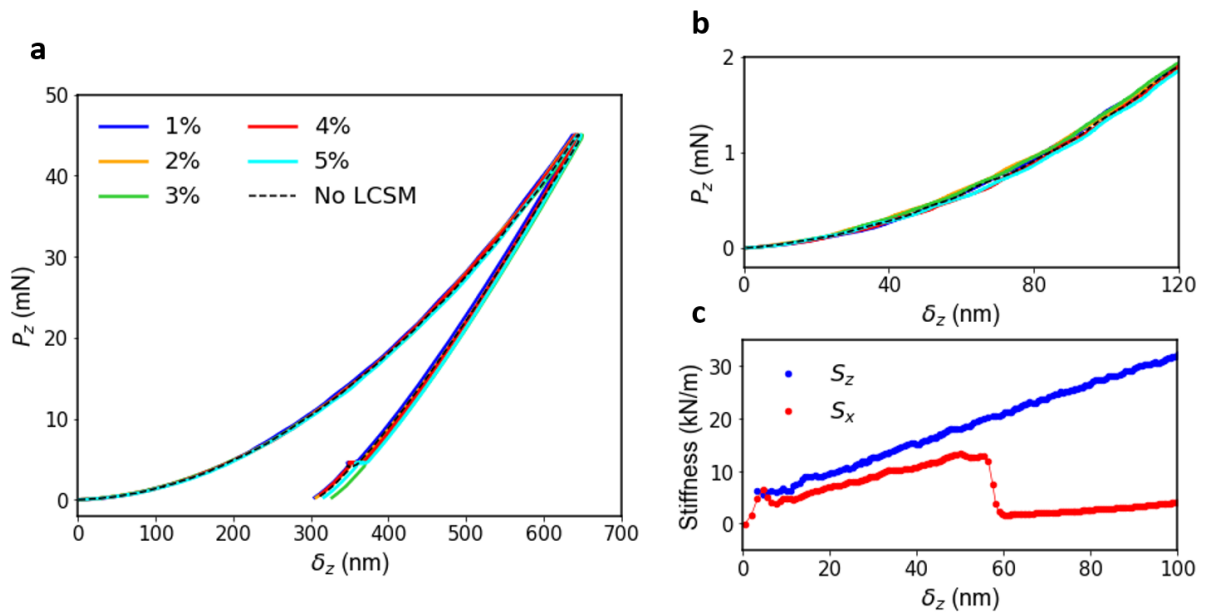

**Figure 5.** (a) Conventional load – displacement curves $P_z$ - $\delta_z$ for the 5 Berkovich LCSM indents into fused silica considered in this section and an indent with no LCSM. (b) Zoomed in view of low load behaviour. Note the stick to slip transition seen in the previous Fig. 4b has no influence on the normal quasi-static behaviour. (c) Normal ($S_z$) and lateral ($S_x$) contact stiffnesses assessed by combined CSM and LCSM measurements on a single constant strain rate test. The stick to slip transition in $S_x$ has no discernible effect on $S_z$

## 4. Lateral plasticity error in single crystal nickel

Many materials of practical interest such as metals and polymers are more prone to ductile yield than the fused silica system studied in the previous section, where brittle fracture is only supressed through the large hydrostatic pressure imposed by the contact geometry.[21] In these materials, it is reasonable to suppose the added shear stress introduced by the LCSM signal may result in additional plastic yield and therefore affect the vertical contact stiffness and load displacement curve. This is the basis of the small amplitude oscillatory shear forming technique developed by *Cross* and co-workers to enhance pattern replication in nanoscale mechanical forming and embossing processes.[50], [51] In the context of parameter extraction through indentation, however, any additional yield on top of what is produced by the mean normal load is obviously undesirable in Poisson's ratio measurements and should be minimized.

In figure 6 we employ the same strategy to investigate the response of a ductile metal to LCSM as was employed for fused silica in the previous section, that is, $\overline{q_x}$ is specified to be a fixed percentage of $P_z$ throughout loading. The metal under investigation is a single crystal nickel <111> surface (Surface Preparation Laboratory, Zaandam, NL) mechanically polished to a mean arithmetic surface roughness of < 5 nm as characterized via AFM. One small change from the previous methodology is that in this case a 1 μm spherical tip replaces the Berkovich geometry. As such, a constant strain rate is not imposed on the sample during loading but rather a constant $\dot{P_z}/P_z = 0.1 \ s^{-1}$.

Fig. 6a shows the vertical load – displacement curves for indentation into nickel with LCSM signals ranging from 1 – 10% of $P_z$. First yield is observable as the jump in depth occurring at roughly 8 nm in all tests. We note this yield point appears in experiments where no LCSM is applied and is generally unaffected by its addition. This is shown in the inset of Fig. 6a, which zooms in on the early quasi-static normal load-displacement behaviour. Included here is an indent with no LCSM signal, plotted in black, which shows plastic yield at approximately the same depth as the LCSM indents. Unlike the fused silica curves of Fig. 5a, the quasi-static behaviour of the nickel beyond the first yield point changes markedly with increasing LCSM amplitude. Less normal load is required to achieve the final depth of $\delta_z = 200$ nm in the 5, 7.5, and 10% tests than in the 1% test. Examining the lateral contact stiffness $S_x$ in Fig. 6b, the behaviour is again distinct from that of fused silica, with no evidence for the dramatic partial-slip to gross slip transition. Rather, at higher $\overline{q_x}$, $S_x$ is seen to decrease. As shown in Fig. 6c, the dramatic increase in $\overline{\delta_x}$ due to slip is also absent in the nickel contact, where the only discontinuity occurs at the point of first yield and otherwise $\overline{\delta_x}$ increases smoothly throughout. The more compliant response both normally and laterally is indicative of additional plastic deformation occurring under tangential loading. Further evidence for this comes in the form of the behaviour of the lateral dynamic phase angle as measured by the lateral lock-in amplifier, which is plotted for the four tests in Fig. 6b. At $\overline{q_x} = 7.5\%$ and $\overline{q_x} = 10\%$ of $P_z$, the phase is noticeably higher than zero throughout indentation. A non-zero CSM phase angle was used as a powerful indicator of increased plasticity in the careful theoretical and experimental study of *Merle* et al. into 'plasticity error' in conventional indentation, where high CSM magnitudes lead to increased plasticity past what is supplied by the mean quasi-static load. We posit that essentially the same behaviour is at work here, albeit with the direction of oscillation shifted 90 degrees.

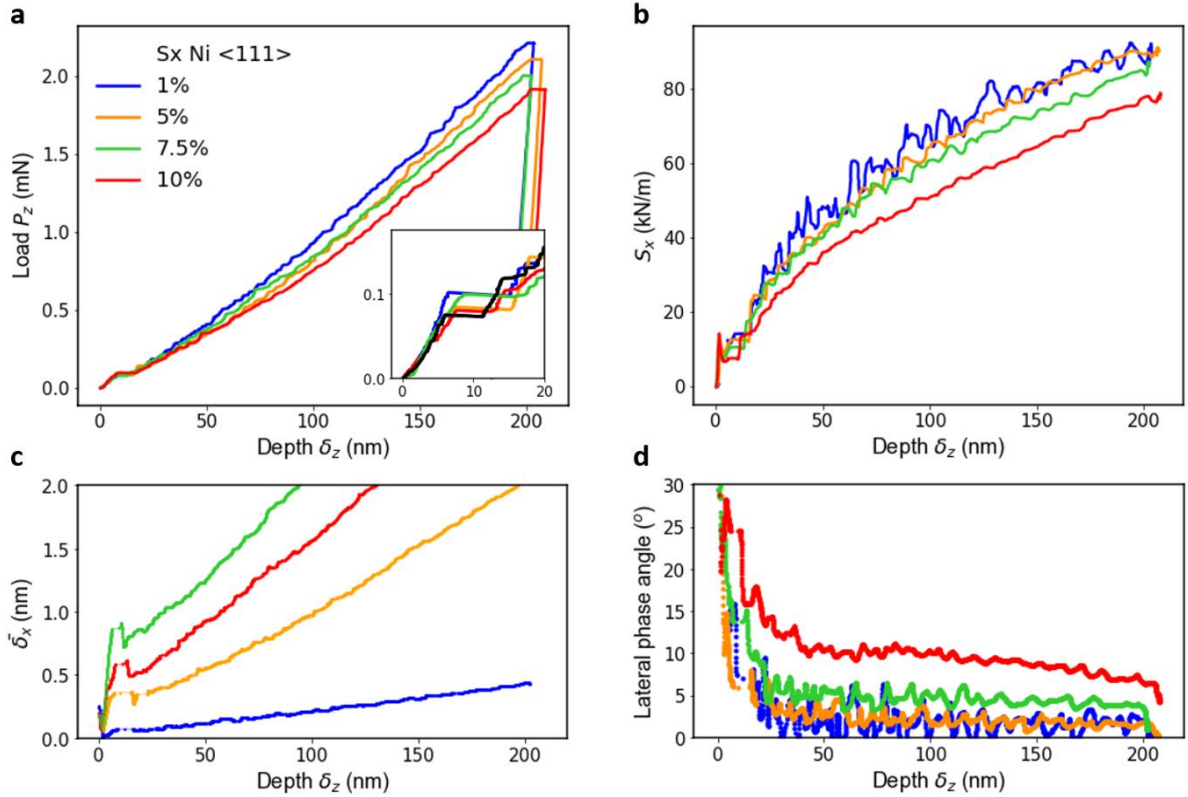

**Figure 6.** (a) Load – displacement curves for indentation into single crystal nickel <111> with a 1 μm radius spherical tip with LCSM loads ranging from 1 – 10% of the normal load $P_z$. (b) Lateral contact stiffness as a function of depth in the same experiments. (c) Lateral oscillation displacement magnitude as a function of depth. (d) Lateral dynamic phase angles during indentation.

## 5. Strategies to limit slip and plasticity in LCSM.

Lateral CSM has the potential to be a powerful tool in repertoire of researchers seeking to characterise the properties of interfaces and the mechanics of small volumes. However, as the previous two sections have demonstrated, significant care must be taken in its deployment. To minimize the influences of slip and added plasticity when measuring quantities like Poisson's ratio, we suggest that two rather simple steps be taken. Firstly, we recommend indenter tip geometries with larger included angles be used. With regard to equation (9), such probes will see the contact area $a$ grow more quickly with respect to the lateral oscillation magnitude than sharper tips, therefore delaying the stick/partial slip to gross slip transition. Secondly, we recommend the loading scheme here, where the LCSM load oscillation $\overline{q_x}$ is specified as a fixed value of $P_z$ throughout indentation, be continued to be used, albeit with some caveats. Evidently, $\overline{q_x}$ must be a small percentage of $P_z$ to prevent the stick-slip transition seen in Fig. 4 or the additional plasticity in Fig. 6b. We suggest 2% as a reasonable compromise in minimizing the noise encountered at very low $\overline{q_x}$ values and avoiding the aforementioned undesirable effects at higher $\overline{q_x}$. The alternative approach of using a fixed (L)CSM displacement amplitude, is presently being phased out in conventional vertical indentation testing due to the critical errors in the observed stiffness that result at low depth as reported by *Pharr* et al.[31] We do not recommend its employment in the lateral case for precisely the same reasons: a fixed amplitude means that at low indentation depths, the LCSM load will be a large fraction of the total load applied to the contact. It can be seen from Fig. 4b that the loads required to initial gross slip in fused silica are around 2 nm, previously the typical fixed CSM displacement used in normal testing. By choosing to apply a fixed $\overline{\delta_x}$ throughout indentation, as is done in the blue curve of Fig. 7, one essentially causes the contact to enter the low stiffness, gross slip state almost immediately. The contact then transitions from the low stiffness state towards a higher $S_x$ stick state as the normal load increases relative to $\overline{\delta_x}$ at greater depths. Hence the fixed $\overline{\delta_x} = 0.7$ nm curve in Fig. 7 re-joining the well-behaved fixed $\overline{q_x} = 2\% \, P_z$ curve (green) at $\delta_z \sim 50 \, nm$. Fixed $\overline{\delta_x}$ values on the order of 1 nm would also result in additional plasticity in metals, at least at low loads, as can be seen in Fig. 6c where $\overline{\delta_x}$ is in the range $1 - 2$ nm to depths of 100 nm. We note that a fixed $\overline{\delta_x}$ has been studied using the prototypical multiaxial indenter mentioned previously.[14] Finally, as noted by *Merle* et al.[33], we suggest that the dynamic lateral phase be monitored during indentation in order to detect significant deviations from the elastic Mindlin solution. As was mentioned in the discussion centred around Fig. 6d, phase angles greater than zero are indicative of additional dissipative processes like slip or plasticity and are therefore a significant indicator that the experiment must be adjusted to ensure accurate results.

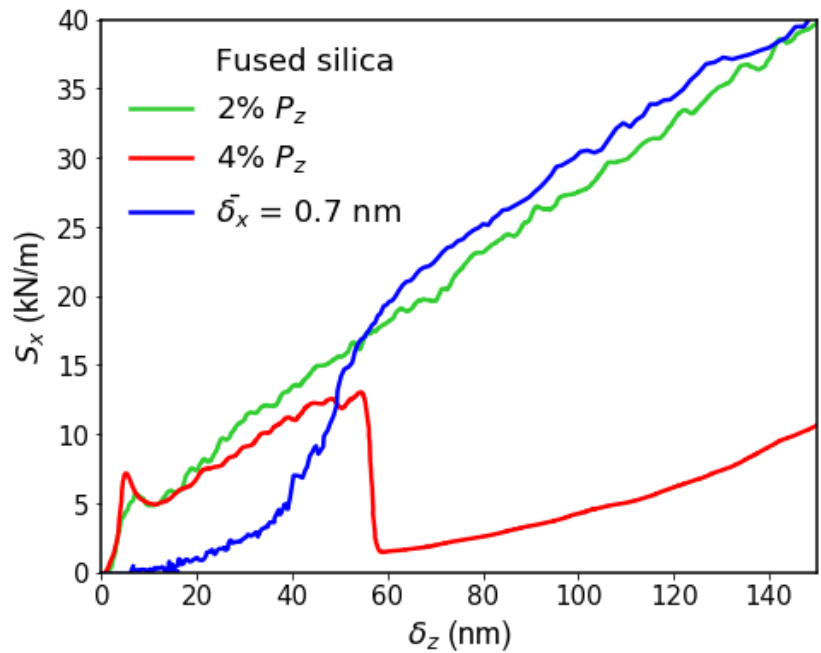

**Figure 7.** Lateral contact stiffness for constant strain rate indentation into fused silica with a Berkovich tip for three LCSM experiments: one with a fixed lateral oscillating displacement magnitude (blue) and two with the lateral oscillating load as a fixed percentage of $P_z$.

## 6. Closing Remarks

Poisson's ratio is a quantity of fundamental scientific importance,[17] describing not only the relative importance of compression and shape change for elastically deformed materials, but also yielding information on atomic packing density and inelastic deformation processes, e.g. densification versus shear flow,[21] and informing the design of modern materials. Poisson's ratio is typically highly sensitive to the structure of the sample in question, with negative (auxetic) values observed in unusually ordered materials such as honeycomb structures, colloidal crystals, and re-entrant polymer foams.[17] Exotic matter structures are generally more common at the nanoscale than in the macroscopic world, e.g., layered 2-D materials and porous low-k dielectric films. Existing methods for measuring Poisson's ratio at small scales, however, typically rely either on exhaustive geometric refinement of the sample in question or intense post-experiment data analysis.[26]–[28], [52] The process described here of applying simultaneous CSM and LCSM oscillations to during a constant strain rate indentation experiment provides a precise and reliable measure of $\nu$ with no additional sample preparation required. To ensure accurate measurement, the roles of dissipative processes such as slip and plasticity must be well-understood and quantifiable to the experimentalist. We have shown that both phenomena may be minimised by correct choice of indenter geometry and LCSM loading protocol. It is hoped the results included here will go some way to inform future researchers on proper design of LCSM experiments to characterise small material volumes.

## Acknowledgments

Support for this work was provided by the Texas A&M President's Excellence Fund X-Grants Program.

# References

[1] W. Oliver, R. Hutchings, and J. Pethica, "Measurement of Hardness at Indentation Depths as Low as 20 Nanometres," in *Microindentation Techniques in Materials Science and Engineering*, 100 Barr Harbor Drive, PO Box C700, West Conshohocken, PA 19428-2959: ASTM International, 1985, pp. 90-90–19.

[2] W. C. Oliver and G. M. Pharr, "An improved technique for determining hardness and elastic modulus using load and displacement sensing indentation experiments," *J. Mater. Res.*, vol. 7, no. 06, pp. 1564–1583, Jun. 1992.

[3] M. D. Uchic, D. M. Dimiduk, J. N. Florando, and W. D. Nix, "Sample dimensions influence strength and crystal plasticity.," *Science*, vol. 305, no. 5686, pp. 986–9, Aug. 2004.

[4] O. Brazil *et al.*, "In situ measurement of bulk modulus and yield response of glassy thin films via confined layer compression," *J. Mater. Res.*, vol. 35, no. 6, pp. 644–653, Mar. 2020.

[5] P. Sudharshan Phani and W. Oliver, "Ultra High Strain Rate Nanoindentation Testing," *Materials (Basel).*, vol. 10, no. 6, p. 663, Jun. 2017.

[6] B. Merle, W. H. Higgins, and G. M. Pharr, "Critical issues in conducting constant strain rate nanoindentation tests at higher strain rates," *J. Mater. Res.*, vol. 34, no. 20, pp. 3495–3503, 2019.

[7] G. M. Pharr, W. C. Oliver, and D. S. Harding, "New evidence for a pressure-induced phase transformation during the indentation of silicon," *J. Mater. Res.*, vol. 6, no. 6, pp. 1129–1130, 1991.

[8] J. . Kim, Y. Choi, S. Suresh, and A. S. Argon, "Nanocrystallization during nanoindentation of a bulk amorphous metal alloy at room temperature," *Science (80-. ).*, vol. 295, no. 5555, pp. 654–657, Mar. 2002.

[9] J. R. Greer and J. T. M. De Hosson, "Plasticity in small-sized metallic systems: Intrinsic versus extrinsic size effect," *Prog. Mater. Sci.*, vol. 56, no. 6, pp. 654–724, Aug. 2011.

[10] J. Hay, P. Agee, and E. Herbert, "Continuous stiffness measurement during instrumented indentation testing.," *Exp. Tech.*, vol. 34, no. 3, pp. 86–94, Jan. 2010.

[11] D. L. Goldsby, A. Rar, G. M. Pharr, and T. E. Tullis, "Nanoindentation creep of quartz, with implications for rate- and state-variable friction laws relevant to earthquake mechanics," *J. Mater. Res.*, vol. 19, no. 1, pp. 357–365, Jan. 2004.

[12] E. G. Herbert, W. C. Oliver, A. Lumsdaine, and G. M. Pharr, "Measuring the constitutive behavior of viscoelastic solids in the time and frequency domain using flat punch nanoindentation," *J. Mater. Res.*, vol. 24, no. 03, pp. 626–637, Mar. 2009.

[13] Y. F. Gao, B. N. Lucas, J. C. Hay, W. C. Oliver, and G. M. Pharr, "Nanoscale incipient asperity sliding and interface micro-slip assessed by the measurement of tangential contact stiffness," *Scr. Mater.*, vol. 55, no. 7, pp. 653–656, Nov. 2006.

[14] J. Annett, Y. Gao, G. L. W. Cross, E. G. Herbert, and B. N. Lucas, "Mesoscale friction

anisotropy revealed by slidingless tests," *J. Mater. Res.*, vol. 26, no. 18, pp. 2373–2378, 2011.

[15] O. Brazil and G. M. Pharr, "Direct observation of partial interface slip in micrometre-scale single asperity contacts," *Tribol. Int.*, p. 106776, Nov. 2020.

[16] B. N. Lucas, J. C. Hay, and W. C. Oliver, "Using multidimensional contact mechanics experiments to measure Poisson's ratio," *J. Mater. Res.*, vol. 19, no. 1, pp. 58–65, Jan. 2004.

[17] G. N. Greaves, A. L. Greer, R. S. Lakes, and T. Rouxel, "Poisson's ratio and modern materials," *Nat. Mater.*, vol. 10, no. 11, pp. 823–837, 2011.

[18] T. Rouxel, "Elastic properties and short-to medium-range order in glasses," *J. Am. Ceram. Soc.*, vol. 90, no. 10, pp. 3019–3039, Oct. 2007.

[19] B. Bridfe and A. A. Higazy, "A model of the compositional dependence of the elastic moduli of polycomponent oxide glasses," *Phys. Chem. Glas.*, vol. 27, no. 1, 1986.

[20] R. E. A. McKnight, T. Moxon, A. Buckley, P. A. Taylor, T. W. Darling, and M. A. Carpenter, "Grain size dependence of elastic anomalies accompanying the α-β Phase transition in polycrystalline quartz," *J. Phys. Condens. Matter*, vol. 20, no. 7, p. 075229, Feb. 2008.

[21] T. Rouxel, H. Ji, J. P. Guin, F. Augereau, and B. Rufflé, "Indentation deformation mechanism in glass: Densification versus shear flow," *J. Appl. Phys.*, vol. 107, no. 9, p. 094903, May 2010.

[22] G. M. Pharr, W. C. Oliver, and F. R. Brotzen, "On the generality of the relationship among contact stiffness, contact area, and elastic modulus during indentation," *J. Mater. Res.*, vol. 7, no. 3, pp. 613–617, 1992.

[23] D. R. Mindlin, "Compliance of Elastic Bodies in Contact," *J. Appl. Mech., ASME*, vol. 16, pp. 259–268, 1949.

[24] B. N. Lucas, J. C. Hay, and W. C. Oliver, "Using multidimensional contact mechanics experiments to measure Poisson's ratio," *J. Mater. Res.*, vol. 19, no. 1, pp. 58–65, 2004.

[25] J. E. Zorzi and C. A. Perottoni, "Estimating Young's modulus and Poisson's ratio by instrumented indentation test," *Mater. Sci. Eng. A*, vol. 574, pp. 25–30, Jul. 2013.

[26] J. H. Kim, S. C. Yeon, Y. K. Jeon, J. G. Kim, and Y. H. Kim, "Nano-indentation method for the measurement of the Poisson's ratio of MEMS thin films," in *Sensors and Actuators, A: Physical*, 2003, vol. 108, no. 1–3, pp. 20–27.

[27] L. Li *et al.*, "Simultaneous determination of the Young's modulus and Poisson's ratio in micro/nano materials," *J. Micromechanics Microengineering*, vol. 19, no. 12, p. 125027, Nov. 2009.

[28] L. Banks-Sills, Y. Hikri, S. Krylov, V. Fourman, Y. Gerson, and H. A. Bruck, "Measurement of Poisson's ratio by means of a direct tension test on micron-sized specimens," *Sensors Actuators, A Phys.*, vol. 169, no. 1, pp. 98–114, Sep. 2011.

[29] X. Li and B. Bhushan, "A review of nanoindentation continuous stiffness measurement technique and its applications," *Mater. Charact.*, vol. 48, no. 1, pp. 11–36, 2002.

[30] K. Durst, B. Backes, O. Franke, and M. Göken, "Indentation size effect in metallic materials: Modeling strength from pop-in to macroscopic hardness using geometrically necessary dislocations," *Acta Mater.*, vol. 54, no. 9, pp. 2547–2555, May 2006.

[31] G. M. Pharr, J. H. Strader, and W. C. Oliver, "Critical issues in making small-depth mechanical property measurements by nanoindentation with continuous stiffness measurement," *J. Mater. Res.*, vol. 24, no. 3, pp. 653–666, Mar. 2009.

[32] S. J. Vachhani, R. D. Doherty, and S. R. Kalidindi, "Effect of the continuous stiffness measurement on the mechanical properties extracted using spherical nanoindentation," *Acta Mater.*, vol. 61, no. 10, pp. 3744–3751, Jun. 2013.

[33] B. Merle, V. Maier-Kiener, and G. M. Pharr, "Influence of modulus-to-hardness ratio and harmonic parameters on continuous stiffness measurement during nanoindentation," *Acta Mater.*, vol. 134, pp. 167–176, Aug. 2017.

[34] P. S. Phani, W. C. Oliver, and G. M. Pharr, "Understanding and modeling plasticity error during nanoindentation with continuous stiffness measurement," *Mater. Des.*, vol. 194, p. 108923, Sep. 2020.

[35] K. L. Johnson, *Contact Mechanics*. Cambridge: Cambridge University Press, 1985.

[36] F. P. Bowden and D. Tabor, *The friction and lubrication of solids*. Oxford: Oxford University Press, 1950.

[37] Y. Mo, K. T. Turner, and I. Szlufarska, "Friction laws at the nanoscale," *Nature*, vol. 457, no. 7233, pp. 1116–1119, Feb. 2009.

[38] B. Luan and M. O. Robbins, "The breakdown of continuum models for mechanical contacts," *Nature*, vol. 435, no. 7044, pp. 929–932, Jun. 2005.

[39] K. L. Johnson, "Surface interaction between elastically loaded bodies under tangential forces," *Proc. R. Soc. London. Ser. A. Math. Phys. Sci.*, vol. 230, no. 1183, pp. 531–548, Jul. 1955.

[40] B. N. J. Persson, "Area of Real Contact: Elastic and Plastic Deformations," in *Sliding Frction: Physical Principles and Applications*, Heidelberg: Springer Berlin Heidelberg, 2000, pp. 45–91.

[41] W. C. Oliver and G. M. Pharr, "Measurement of hardness and elastic modulus by instrumented indentation: Advances in understanding and refinements to methodology," *J. Mater. Res.*, vol. 19, no. 01, pp. 3–20, Jan. 2004.

[42] J. B. Pethica and W. C. Oliver, "Mechanical Properties of Nanometre Volumes of Material: use of the Elastic Response of Small Area Indentations," *MRS Proc.*, vol. 130, 1988.

[43] O. Brazil and G. M. Pharr, "Direct observation of partial slip in micrometre-scale single asperity contacts," *Tribol. Int. - Accept.*, 2020.

[44] I. N. Sneddon, "The relation between load and penetration in the axisymmetric boussinesq problem for a punch of arbitrary profile," *Int. J. Eng. Sci.*, vol. 3, no. 1, pp. 47–57, May 1965.

[45] J. C. Hay, A. Bolshakov, and G. M. Pharr, "Critical examination of the fundamental relations used in the analysis of nanoindentation data," *J. Mater. Res.*, vol. 14, no. 6,

pp. 2296–2305, 1999.

[46] S. I. Bulychev, V. P. Alekhin, M. K. Shorshorov, and A. P. Ternorskii, "Determining Young's modulus from the indenter penetration diagram," *Zavod. Lab*, vol. 41, no. 9, pp. 11137–11140, 1975.

[47] E. Harsono, S. Swaddiwudhipong, and Z. S. Liu, "The effect of friction on indentation test results," *Model. Simul. Mater. Sci. Eng.*, vol. 16, no. 6, p. 065001, Sep. 2008.

[48] W. C. Guo, G. Rauchs, W. H. Zhang, and J. P. Ponthot, "Influence of friction in material characterization in microindentation measurement," in *Journal of Computational and Applied Mathematics*, 2010, vol. 234, no. 7, pp. 2183–2192.

[49] M. Mata and J. Alcalá, "The role of friction on sharp indentation," *J. Mech. Phys. Solids*, vol. 52, no. 1, pp. 145–165, Jan. 2004.

[50] G. L. W. Cross, B. S. O'Connell, H. O. Özer, and J. B. Pethica, "Room temperature mechanical thinning and imprinting of solid films," *Nano Lett.*, vol. 7, no. 2, pp. 357–362, 2007.

[51] O. Brazil, V. Usov, J. B. Pethica, and G. L. W. Cross, "Large area thermal nanoimprint below the glass transition temperature via small amplitude oscillatory shear forming," *Microelectron. Eng.*, vol. 182, 2017.

[52] B. B. Jung, S. H. Ko, H. K. Lee, and H. C. Park, "Measurement of Young's modulus and poisson's ratio of thin film by combination of bulge test and nano-indentation," in *Advanced Materials Research*, 2008, vol. 33-37 PART 2, pp. 969–974.